\documentclass[aps,pra,preprint,groupedaddress,showpacs]{revtex4-1}  
\usepackage{graphicx}  
\usepackage{subfigure} 
\usepackage{setspace}
\usepackage{bm}        
\usepackage{amssymb}   
\usepackage{amsmath}   
\usepackage{url}       
\usepackage[colorlinks,linkcolor=blue,anchorcolor=blue,citecolor=blue,urlcolor=black]%
{hyperref}
\begin{document}
\title{Dissipative optomechanics of a single-layer graphene in a microcavity}
\author{Lin-Da Xiao$^1$}
\author{Yu-Feng Shen$^1$}
\author{Yong-Chun Liu$^{1,2}$}
\email{ycliu@pku.edu.cn;}
\author{Meng-Yuan Yan$^1$}
\author{Yun-Feng Xiao$^{1,2}$}
\altaffiliation{url:www.phy.pku.edu.cn/$\sim$yfxiao/index.html}
\affiliation{$^{1}$State Key Laboratory for Mesoscopic Physics and School of Physics,
Peking University, Beijing 100871, P. R. China}
\affiliation{$^{2}$Collaborative Innovation Center of Quantum Matter, Beijing 100871, P. R. China}
\date{\today}
\begin{abstract}
\indent{}We study the optomechanical coupling of a single-layer graphene with a high-\textit{Q} Fabry-P\'erot microcavity in the ``membrane-in-the-middle'' configuration. In ordinary dissipative coupling systems, mechanical
 oscillators modulate the loss associated with the input coupling of the cavity mode; while in our system, the graphene oscillator couples dissipatively with the cavity mode through modulating its absorption loss. By
 analyzing the effects of the interband transition of a graphene suspended near the node of the cavity field, we obtain strong and tunable dissipative coupling without excessively reducing the optical quality factor. Finally,
 it is found that the flexural mode of the graphene could be cooled down to its ground state in the present coupling system. This study provides new insights for graphene optomechanics in the visible range.
\end{abstract}
\maketitle{}
\section{Introduction}
\indent{}In recent years, optomechanics has attracted much attention,
  since it not only provides an excellent platform for fundamental test of quantum theory and exploration of quantum-classical boundary, but also has important applications in quantum information processing and precision metrology \cite{Kippenberg29082008,aspelmeyer2013cavity,Meystre2013short,Liu2014review}.
 So far, many different optomechanical systems have been proposed and explored, and the coupling mechanisms generally can be divided into two groups, dispersive coupling and dissipative coupling. The former modulates the resonance frequency of the cavity \cite{Law1995interaction} while the later varies the damping rate of the cavity \cite{PhysRevLett.102.207209}.
 On the one hand, intensive studies have been conducted on dispersive coupling, resulting in many important consequences. For example, ground-state cooling of the mechanical mode has been predicted \cite{wilson2007theory,PhysRevLett.99.093902,genes2008ground} and demonstrated experimentally \cite{chan2011laser,teufel2011sideband}.
 On the other hand, in dissipative coupling regime, recent theoretical work have also revealed the that mechanical mode can be cooled down to the ground state without the resolved sideband condition \cite{PhysRevLett.102.207209,PhysRevA.88.023850}.
 Such dissipative coupling has been verified experimentally \cite{PhysRevLett.103.223901} and studied in several different systems \cite{xuereb2011dissipative,yan2013dissipative}. It has also shown high potential to generate and manipulate the quantum states of both light \cite{kronwald2014dissipative} and mechanical mode \cite{gu2013generation,kronwald2013arbitrarily}.\\
\indent{}To construct a practical dissipative optomechanical system, it is essential to obtain large enough dissipative coupling and to maintain  plausible mechanical characteristic simultaneously. In this paper, we introduce graphene membrane as a mechanical resonator suspended in a Fabry-P\'erot (FP) cavity.
In previous dispersive ``membrane-in-the-middle'' approach \cite{jayich2008dispersive,thompson2008strong}, membranes possess high reflectivity and low absorbance, thus the dispersive coupling is the dominant regime.
In the present proposal, however, due to the unique 2D hexagonal lattice and Dirac cone band structure, graphene possesses high and frequency-independent absorbance in the visible range \cite{neto2009electronic}, and this strong absorption could be used to generate strong dissipative coupling.
Moreover the graphene mechanical resonator possesses high mechanical \textit{Q}, high resonance frequency and low effective mass \cite{Jiang2014review}.
Very recently, there has been some progress in the research of graphene optomechanics in microwave range \cite{singh2014optomechanical,song2014graphene}, where optomechanically induced transparency and backaction cooling have been observed experimentally,
and they all make use of the conventional dispersive coupling. Here, we focus on dissipative-coupling graphene optomechanics in the visible range.\\
\indent{}The rest of this paper is organized as follows. In section~\ref{sec:model}, we employ a simplified model to analyse the coupling system using the input-output formalism. In section~\ref{sec:coupling}, we investigate the dissipative optomechanical coupling and discuss the cooling of the graphene resonator.
 A short summary is given in section~\ref{sec:conclusion}.
\section{Model}
\label{sec:model}
\label{sec:model}
\begin{figure}
\centering
\includegraphics[width=0.45\textwidth]{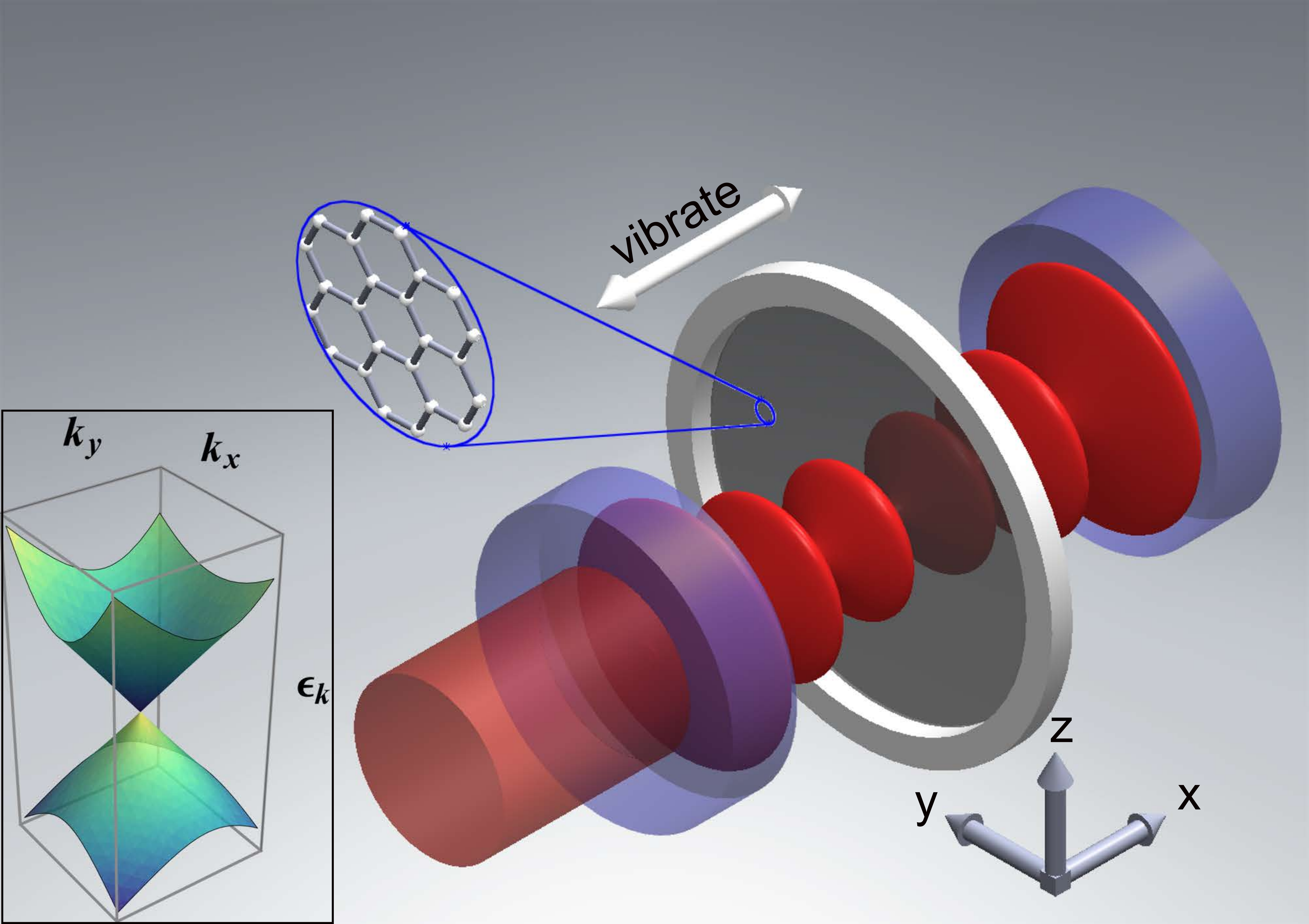}
\caption{\small (Color online) Schematic illustration of the optomechanical system. A graphene film of diameter $D$ is suspended with a fixed frame, and we would consider the first flexural mode vibrating in the direction of $x$ axis.
Inset: Local band structure of graphene at the vicinity of a Dirac point. The parameters of the apparatus are [$L$, $\lambda_\text{cav}$, $\kappa_{\text{c}}$]=[30 $\mu\text{m}$, $600$ nm, $2\pi\times1$ MHz], where $L$, $\lambda_\text{cav}$ and $\kappa_{\text{c}}$ are the effective cavity length, resonant wavelength and linewidth of the cavity mode.}\label{fig:apparatus&band}
\end{figure}

\indent{}Figure.~\ref{fig:apparatus&band} shows the present optomechanical system consisting of a FP cavity and a single layer graphene, which is generally referred to as ``membrane-in-the-middle'' coupling.
Note that the cavity decay includes two contributions, the decay associated with input coupling and the decay not relevant with the input coupling.
Most of the previous dissipative optomechanical systems utilized coupling decay \cite{xuereb2011dissipative,kronwald2014dissipative,gu2013generation,kronwald2013arbitrarily}, while we make use of the absorptive part.
This difference would modify the conventional dissipative coupling regime, and the quantum noise of the inner degree of freedom of graphene would also be involved as an input.

\begin{figure}
\centering{}
\includegraphics[width=0.45\textwidth]{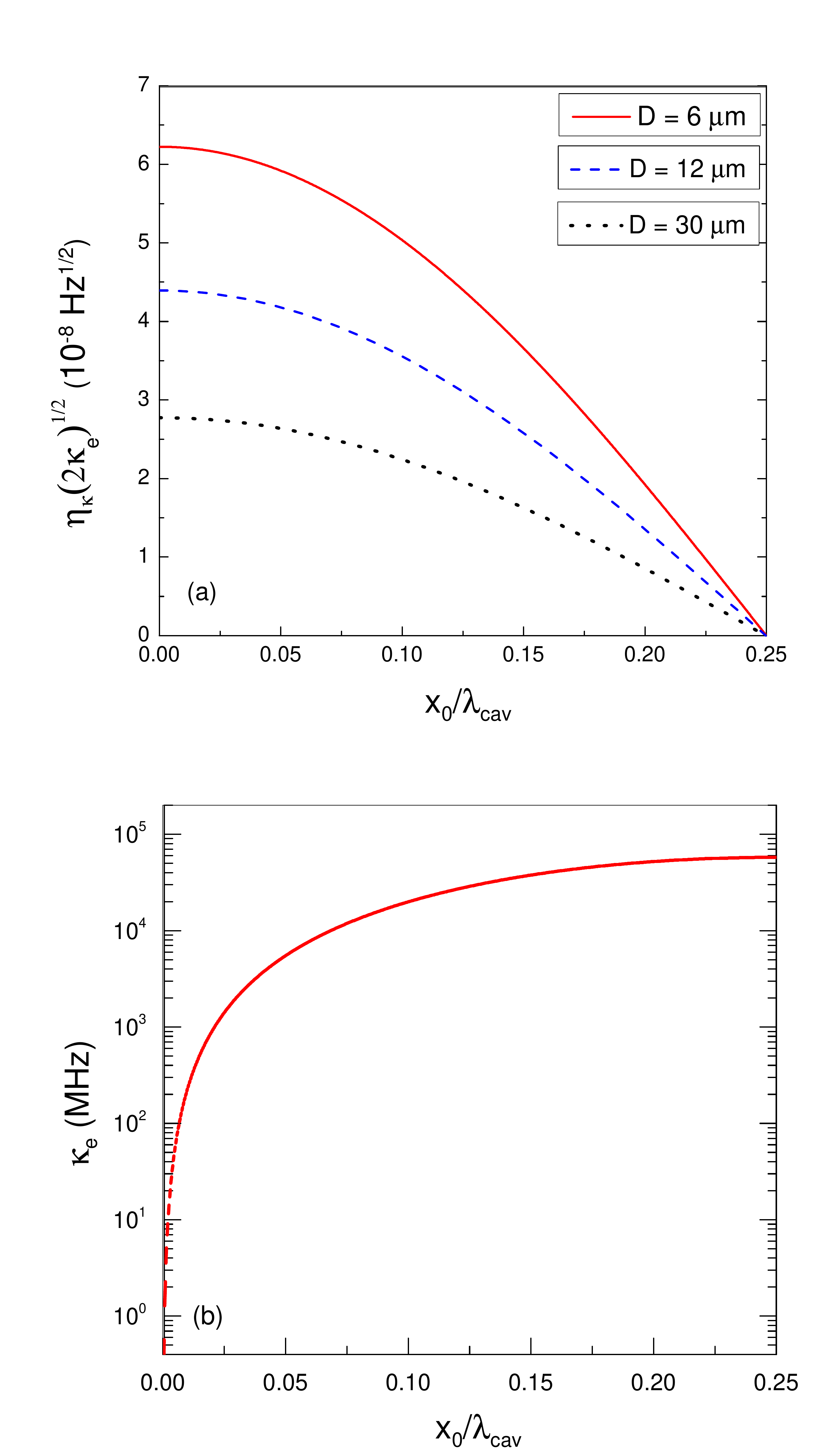}
\caption{\small (Color online) (a) Coupling parameter $\eta_{\kappa}(2\kappa_{\text{e}})^{1/2}=\frac{d}{dx} \sqrt{2 \kappa_{\text{e}}} X_{\text{ZPF}}$ as a function of $x_0$, where $x_0$ is the equilibrium position of mechanical oscillator. Here $x_0=0$ corresponds to a node of a standing-wave cavity field.  (b) Absorptive damping rate $\kappa_{\text{e}}$ as a function of equilibrium position $x_0$.}\label{fig:coupling&damping}
\end{figure}

The total Hamiltonian is written as
\begin{subequations}
\begin{align}
 &H=H_{\text{cav}}+H_{\text{\text{mec}}}+H_{\text{pump}}+H_{\text{e}}+H_{\text{i}},\\
 &H_{\text{cav}}=\hbar\omega_{\text{cav}} a^{\dagger}a,\\
 &H_{\text{mec}}=\hbar\omega_{\text{m}} b^{\dagger}b,\\
 &H_{\text{pump}}=i\hbar(E e^{-i\omega_{\text{p}}t}a^{\dagger}-E^{\ast}e^{i\omega_{\text{p}}t}a),
\end{align}
\end{subequations}
 where $H_{\text{cav}}$, $H_{\text{mec}}$ and $H_{\text{pump}}$ are the ordinary Hamiltonians for cavity field, mechanical resonator and the pump field  while $H_{\text{e}}$ and $H_{\text{i}}$ describe the dynamics of $\pi$ electrons of graphene and the interaction of the optomechanical system. In these Hamiltonians, $\omega_{\text{cav}}$, $\omega_{\text{m}}$ and $\omega_{\text{p}}$ are frequencies of cavity field, mechanical resonator and pumping laser field.
 $a$ is the annihilation operator for cavity field and $b$ for mechanical resonator.
The amplitude $E$ is related to pump power $P$ via the relation $|E|=\sqrt{2 P \kappa_\text{c}/\hbar\omega_\text{p}}$.\\
\indent{}Before getting down to $H_{\text{e}}$ and $H_{\text{i}}$, we digress for a moment to discuss the mechanical properties of graphene oscillator.
Consider a suspended, circular shaped graphene. The mechanical flexural modes have adjustable frequencies, as predicted by the classical circular membrane model \cite{courant1966methods}, the frequency of the first radial symmetric mode is $\omega_{\text{m}}=2\pi\times0.766{(\tilde{E}\xi/\rho)}^{1/2}/D$, where the two dimensional elastic stiffness $\tilde{E}$ is $340\ \text{N/m}$,
corresponding to a Young¡¯s modulus $E$ of graphene approximately 1 TPa, $\rho=7.4 \times{10^{-19}}\ \textrm{kg}/{\textrm{$\mu$m}^{2}}$  (Ref.~\cite{chen2013graphene}), $\xi$ is the strain in graphene sheet which varies from 0.01\% to 10\% according to current experiments \cite{lee2008measurement}, and $D$ is the diameter of the circular graphene.
In the following discussion, we choose $\xi=1\%$ while leave $D$ an adjustable parameter.
\\
\indent{}We then turn to the electron part of the Hamiltonian $H_{\text{e}}$. The band structure of graphene is shown in the inset of Fig.~\ref{fig:apparatus&band}.
At the vicinity of the Dirac point, the energy of single electron can be approximated by $E_\pm(\textbf{q})=\pm\hbar v_\text{F}|\textbf{q}|$. Under this Dirac cone approximation, the general tight-binding Hamiltonian \cite{mahan2010condensed} of the $\pi$ electrons in graphene can be reduced to
\begin{equation}
\begin{aligned}
H_\text{e}=&\underset{\textbf{q}}{\sum}\left(\langle\chi_{\text{c},\textbf{q}}|C_{\text{c},\textbf{q}}^{\dagger}+\langle\chi_{\text{v},\textbf{q}}|C_{\text{v},\textbf{q}}^{\dagger}\right)\times\left(\hbar v_\text{F} \mathbf{\sigma} \cdot \textbf{q}\right)\\
&\times\left(|\chi_{\text{c},\textbf{q}}\rangle C_{\text{c},\textbf{q}}+|\chi_{\text{v},\textbf{q}}\rangle C_{\text{v},\textbf{q}}\right),
\end{aligned}
\end{equation}
where $\textbf{q}$ sums over all allowed wave vector in the reciprocal lattice space, $C_{(\text{c,v}),\textbf{q}}$ and $C_{(\text{c,v}),\textbf{q}}^{\dagger}$ are annihilation and creation operators of the conduction band and valance band, $v_\text{F}$ is the fermi velocity, $\mathbf{\sigma}=({\sigma}_1,{\sigma}_2)$ is the two dimensional form of Pauli matrices.
The spin eigen vector for a given $\textbf{q}$ is $\left|\chi_{(\text{c,v}),\textbf{q}}\right\rangle=1/\sqrt{2}\left(e^{-i\theta(\textbf{q})/2},\pm e^{i\theta(\textbf{q})/2}\right)^{\tiny{\text{T}}}$, where $\text{c,v}$ and ¡°$\pm$¡± denote the conduction and valance band respectively, and $\theta(\textbf{q})=\arccos({q_z}/{q_y})$.\\
\indent{}The interaction is introduced via the minimum coupling, $\hbar\textbf{q}\to \hbar\textbf{q}+e\textbf{A}/c$.
In the simplest case, we consider a single mode field polarizing in the \textit{z} direction.
Implementing the quantized cavity field, the interacting Hamiltonian $H_{\text{i}}$ is,
\begin{equation}\label{graphene_hamiltonian}
\begin{split}
H_{\text{i}}=
&
\underset{\textbf{q}}{\sum}\eta_{\textbf{q}}^{\dagger}\sigma_1 \eta_{\textbf{q}}
A_0 \sin\left(\frac{\omega_\text{cav}}{c} x\right)(a^{\dagger}+a),\\
\end{split}
\end{equation}
where $\eta_{\textbf{q}}^{\dagger}=\langle\chi_{\text{c},\textbf{q}}|C_{\text{c},\textbf{q}}^{\dagger}+\langle\chi_{\text{v},\textbf{q}}|C_{\text{v},\textbf{q}}^{\dagger}$ is a spinor operator, $\sigma_1$ is a Puali matrix, $A_0=\sqrt{\hbar/{\epsilon_0 V \omega_{\text{cav}}}}$ and $x$ is the position of the graphene referring to the cavity.\\
\indent{}We follow the approach of Ref.~\cite{mecklenburg2010tree} to calculate the energy absorption of normal incident light per unit area using Fermi's golden rule, and we can recover the well-known universal absorption in visible range \cite{nair2008fine},
$
W=\pi \alpha c[f(E_\text{c})-f(E_\text{v})]u(\mathbf{r}),
$
where 
$\alpha$ is the fine structure constant, $u(\mathbf{r})$ is the electro-magnetic energy density and $f(E)$ is the Fermi-Dirac distribution function. Under the approximation of low temperature, the result reduces to $\pi\alpha c u(\mathbf{r})$.
That is, for clean, undoped graphene, about 2.3\% of perpendicular incident light is dissipated via the spontaneous emission channel.
Then we can subsequently obtain the expression of damping rate of the cavity caused by dynamics of electrons
\begin{equation}
\kappa_{e}=\frac{{W_{\text{max}}} A_{\text{eff}}}{\int u\textit{d}V}\label{kappa}.
\end{equation}
Here, $A_{\text{eff}}=\int u d y dz/u_{\text{max}}$ is the effective area of graphene as a function of the longitudinal position $x$, and it is defined to be an analogy of mode volume. We obtain $W_{\text{max}}(x)$ and $u_{\text{max}}(x)$ at the center of the beam.
We approximate the field distribution by Gaussian beam, and the numerical simulation is plotted in Fig.~\ref{fig:coupling&damping}. Neglecting other minor channels of dissipation, in our system, $\kappa_{\text{e}}$ dominates the dissipative coupling.
\\
\indent{}For convenience, we use a set of Pauli matrixes to describe the transition between the conduction band and valence band, $C_{\text{c},{\textbf{q}}}^{\dagger}C_{\text{v},\textbf{q}}\rightarrow \sigma_{\textbf{q}}^{\dagger}$, $C_{\text{v},{\textbf{q}}}^{\dagger}C_{\text{c},\textbf{q}}\rightarrow \sigma_{\textbf{q}}$, and $1/2(C_{\text{c},{\textbf{q}}}^{\dagger}C_{\text{c},\textbf{q}}-C_{\text{v},{\textbf{q}}}^{\dagger}C_{\text{v},\textbf{q}})\rightarrow \sigma_{\textbf{q}}^{z}$.
Then the system is similar to cavity QED system of multiple two-level atoms interacting with a single cavity mode, and the techniques in cavity QED is expected be applied here. In the input-output formalism, we define
$\sigma_{\text{e}}=\sum_{\textbf{q}}'g_{\textbf{q}}\sigma_{\textbf{q}}\approx\sum_{\textbf{q}}'B\sigma_{\textbf{q}}$,
where $g_{\textbf{q}}$ is the coupling constant of ``two level atom'' with light field and constant $B$ comes from the details of input-output formalism. $B$ is introduced to make the dimension of $\sigma_\text{e}$ correspond to a standard input noise operator and $\sum_{\textbf{q}}'$ means the summation is restricted to those states of
the energy $\pm\hbar\omega_{\text{cav}}/2$.
Inserting these back into Eq.~\ref{graphene_hamiltonian}, dropping the high frequency terms in the rotating wave approximation, we can obtain the familiar form,
$
H_{\text{e}}+H_{\text{i}}=
{\sum}_{\textbf{q}}\hbar v_\text{F}|q| \sigma_{\textbf{q}}^{z}+
i\hbar\sqrt{2\kappa_{\text{e}}}
(\sigma_{\text{e}}^{\dagger}a-\sigma_{\text{e}}a^{\dagger}).
$
Because of the high relaxation rate of electrons in graphene, the field is often not strong enough to appreciably influence the occupation of electrons, in other word, $\langle\sigma_{\textbf{q}}^z\rangle=-1/2$. This means we can drop the $H_{\text{e}}$ now, for it only contributes a constant to the ground state energy in our system.
In the input-output formalism, the coupling constant $g_{\textbf{q}}$ of ``two level atom'' with light field is required to be a slow-varying function of $\textbf{q}$ \cite{walls2007quantum}, which is not the case in our system.
But for a first order calculation only, $\sum_{\textbf{q}}' g_{\textbf{q}}\sigma_{\textbf{q}}$ is expected to be replaced with an average.
The expectation value of the operator $\sigma_\text{e}$ can be obtained by the Kubo formula \cite{mahan2000many},
\begin{equation}\label{linear_response}
\begin{split}
\langle {{\sigma}_\text{e}}(t)\rangle
&\approx\langle {{\sigma}_\text{e}}(0)\rangle+\frac{1}{i \hbar} \int_{-\infty}^{\infty}\theta(t-t')\left\langle [{{\sigma}_\text{e}}(t) , {H}_{\text{i}}(t')]\right\rangle_0 dt'\\
&\approx-\frac{1}{8}\sqrt{2\kappa_e}\frac{{v_\text{F}}^2}{c^2}\frac{A_{\text{eff}}}{A}|\bar{a}|^2e^{-i \omega t},
\end{split}
\end{equation}
where $A$ is the total area of graphene, $\bar{a}$ is the expectation of the cavity filed operator $a$ and $\theta(t-t')$ is a unit step function. For graphene, the ratio of $v_\text{F}/c$ is about 1/300, which guarantees the validity of Kubo formula in linear response regime.\\
\section{Dissipative Coupling of the Graphene}
\label{sec:coupling}
\begin{figure}
\centering{}
\includegraphics[width=8cm]{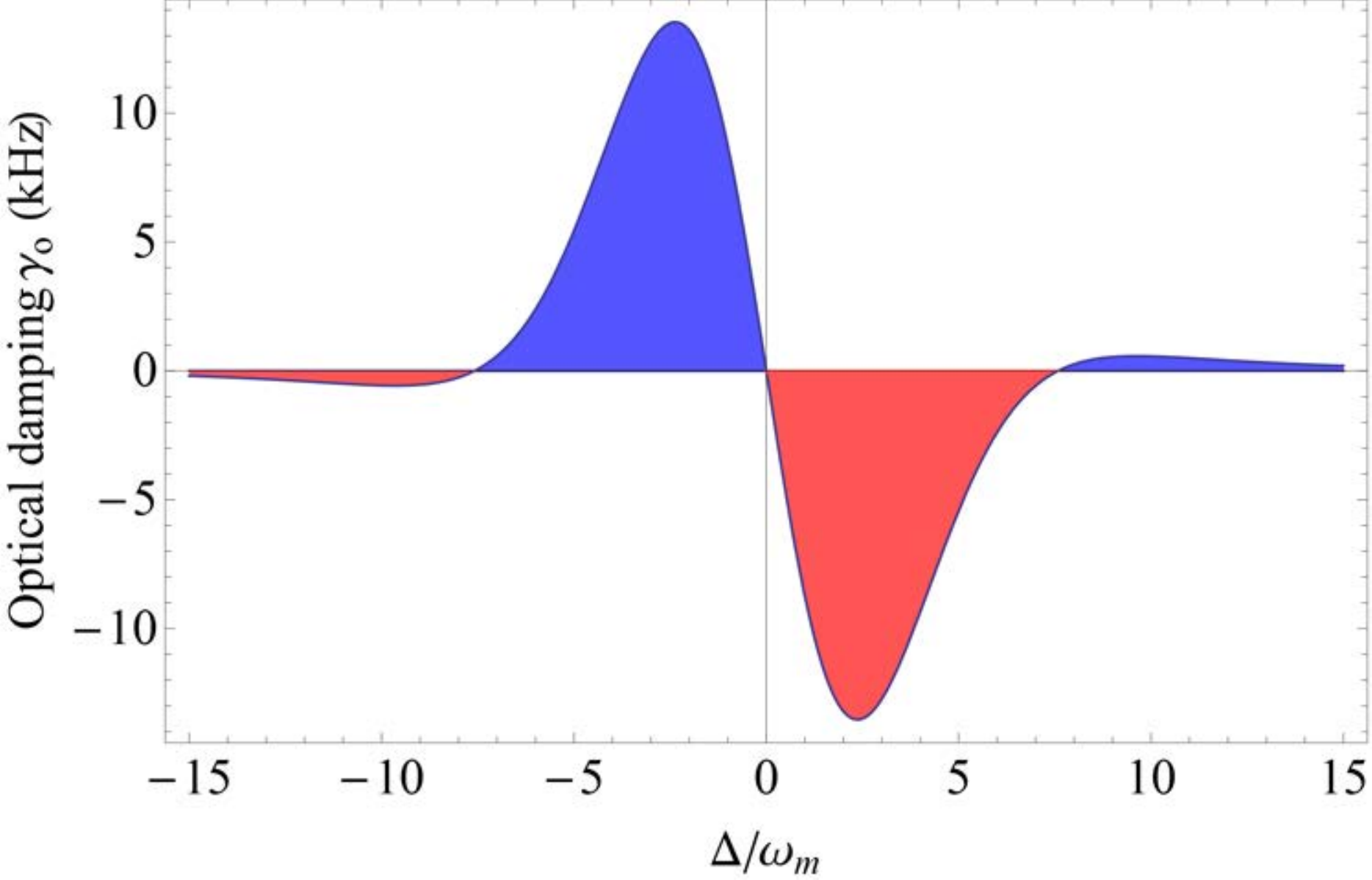}
\caption{\small (Color Online) The optical induced damping as a function of detuning $\Delta$, where the blue and red area correspond to the cooling and heating region, respectively. The pump power $P=5\ \mu \text{W}$, and other parameters are $D=30\ \mu\text{m}, A_{\text{eff}}/A=0.01, \omega_{\text{m}}=2\pi\times 55\ \text{MHz}, \kappa_{\text{c}}=2\pi\times1 \ \text{MHz},\kappa_e=2\pi\times45\ \text{MHz}, \eta_{\kappa}=2.2\times 10^{-3}$. }\label{fig:damping}
\end{figure}

\begin{figure}
\centering{}
\includegraphics[width=10cm]{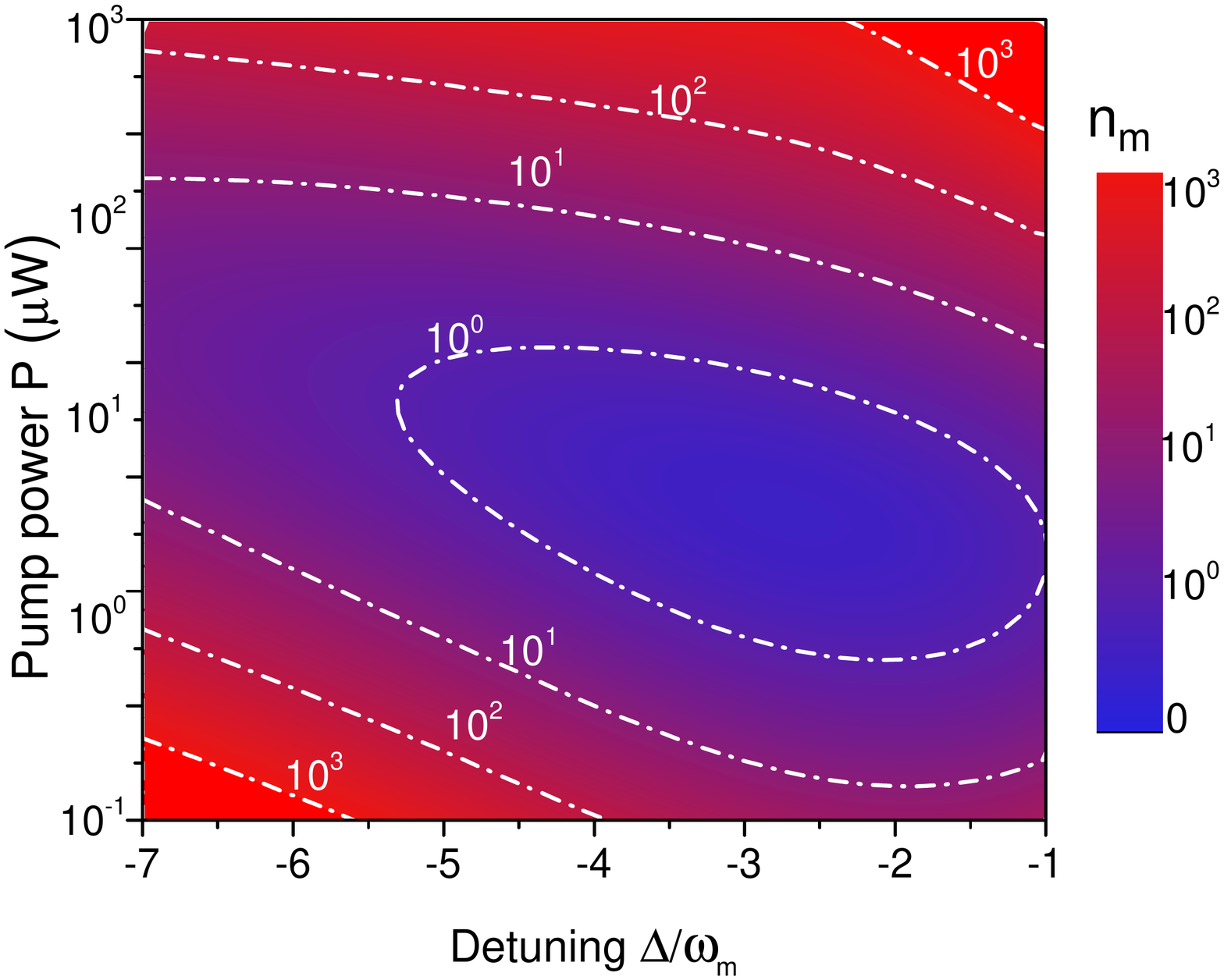}
\caption{\small (Color online) The steady state phonon number, based on Eq.~\ref{phonon_number}. Steady state phonon number is illustrated with the contour and color.}\label{fig:phonon}
\end{figure}

\indent{}Expanding the absorptive damping around the position of equilibrium, we can obtain the total Hamiltonian.
\begin{equation}
\begin{split}
H=&\hbar\omega_{\text{cav}}a^{\dagger}a+\hbar\omega_{\text{m}}b^{\dagger}b+i\hbar(E e^{-i\omega_{\text{p}}t}a^{\dagger}-E^{\ast}e^{i\omega_{\text{p}}t}a)\\
&+i\hbar\sqrt{2\kappa_{\text{e}}}({\sigma}_{\text{e}}a^{\dagger}-{{\sigma}^{\dagger}_{\text{e}}}a)[1+\eta_{\kappa}(b^{\dagger}+b)].
\end{split}
\end{equation}
The last term introduces the dissipative optomechanical coupling, where
 \begin{equation}
 \eta_{\kappa}=\frac{d}{dx} \sqrt{2 \kappa_{\text{e}}(x)} X_{\text{ZPF}}/\sqrt{2\kappa_{\text{e}}},
\end{equation}
 and $X_{\text{ZPF}}=\sqrt{\hbar/2 m_{\text{eff}} \omega_{\text{m}}}$. For a circular membrane, the effective mass is approximately 0.27 times the mass of membrane. In the case the coupling strength depends on the position and diameter of the graphene, as is shown in Fig.~\ref{fig:coupling&damping}.
We find that the reliance of $\kappa_{\text{e}}$ on equilibrium position $x_0$ mimics the distribution of intracavity field, and smaller diameter can increase $X_{\text{ZPF}}$ thus to generate stronger coupling.\\
\indent{}We then explore this system via the standard quantum noise approach. First, we write the Langevin equation in rotating frame of the pump field
\begin{subequations}
\begin{align}
&\dot{a}=-\left[i\Delta+\kappa+2\kappa_{\text{e}}\eta_{\kappa}(b^{\dagger}+b)\right]a+E+\sqrt{2\kappa_{\text{c}}}a_{\text{in}}\nonumber\\
&\ \ \ \ \ +\sqrt{2\kappa_{\text{e}}}{\sigma}_{\text{e}}[1+\eta_{\kappa}(b^{\dagger}+b)],\\
&\dot{b}=-(i\omega_{\text{m}}+\gamma_{\text{m}})b+\sqrt{2\gamma_{\text{m}}}b_{\text{in}}(t)+\sqrt{2\kappa_{\text{e}}}\eta_{\kappa}({\sigma}_{\text{e}}a^{\dagger}-{\sigma}^{\dagger}_{\text{e}}a),
\end{align}
\end{subequations}
where $\Delta=\omega_{\text{cav}}-\omega_{\text{p}}$ is the detuning. $b_{\text{in}}$ is the thermal noise of the mechanical resonator while $a_{\text{in}}$ is the noise operator of light.
Their  two-frequency-correlation functions are $\langle a_{\text{in}}(\omega)a_{\text{in}}(\omega')\rangle=0$, $\langle b_{\text{in}}(\omega)b_{\text{in}}(\omega')\rangle=0$, $\langle a^{\dagger}_{\text{in}}(\omega)a_{\text{in}}(\omega')\rangle=\delta(\omega+\omega')$ and $\langle b^{\dagger}_{\text{in}}(\omega)b_{\text{in}}(\omega')\rangle=n_{\text{th}}\delta(\omega+\omega')$ where $n_{\text{th}}$ is the thermal thermal phonon number. Then the steady state solution of intracavity field is
\begin{equation}\label{average_field}
\bar{a}=\frac{E+\sqrt{2\kappa_{\text{e}}}\bar{\sigma}_{\text{e}}}{i\Delta+\kappa_\text{c}+\kappa_\text{e}}.
\end{equation}
 Equation~\ref{linear_response} combined with Eq.~\ref{average_field} can be solved self-consistently, and both $\bar{a}$ and $\bar{\sigma}_\text{e}$ would eventually only rely on the input $E$. Now, we can expand the operators at the mean value: $a=\bar{a}+\delta{a}$ and ${\sigma}_{\text{e}}=\bar{\sigma}_{\text{e}}+{\sigma}_{\text{in}}$ and linearize the equation. Here, $\sigma_{e}$ serves as an intermediate field and it introduce additional quantum noise. Transforming these linearized equations to frequency domain, i.e., $a(\omega)=\int dt e^{-i\omega t} \delta a(t)$, we obtain\\
 \begin{subequations}
 \begin{align}
&(-i\omega+i\Delta+\kappa)a(\omega)=-G\left[b^{\dagger}(\omega)+b(\omega)\right]+\sqrt{2\kappa_\text{c}}a_{\text{in}}(\omega)\nonumber\\
&\ \ \ \ \ \ \ \ \ \ \ \ \ \ \ \ \ \ \ \ \ \ \ \  \ \ \ \ \ \ +\sqrt{2\kappa_\text{e}}\sigma_{\text{in}}(\omega)\\
&(-i\omega+i\omega_{\text{m}}+\gamma_{\text{m}})b(\omega)=\sqrt{2\gamma_{\text{m}}}b_{\text{in}}(\omega)+X_{\text{ZPF}} F(\omega)\\
&F(\omega)=\frac{\sqrt{2\kappa_{\text{e}}}}{X_{\text{ZPF}}}\left[\frac{G_{\text{e}}}{\sqrt{2\kappa_{\text{e}}}} a^{\dagger}(\omega)+\sigma_{\text{in}}(\omega)\bar{a}^{\ast}-\text{h.c.}\right].
\end{align}
\end{subequations}
Here, $G$ and $G_{\text{e}}$ are shorthand notation of $(2 \bar{a}\kappa_{\text{e}}-\sqrt{2\kappa_{\text{e}}}\bar{\sigma}_{\text{e}})\eta_{\kappa}$ and $\sqrt{2\kappa_\text{e}}\bar{\sigma}_{\text{e}}\eta_{\kappa}$, respectively.
In the weak coupling regime, i.e., $\left|G\right|,\left|G_\text{e}\right|\ll \kappa$, we can obtain an approximate solution for $b(\omega)$
\begin{widetext}
\begin{equation}\label{approximate_solution}
\begin{split}
b(\omega)=&\frac{\sqrt{2\gamma_{\text{m}}}b_{\text{in}}(\omega)}{-i\omega+i\tilde{\omega}_{\text{m}}+\tilde{\gamma}_{\text{m}}}+\frac{G\left[\sigma_{\text{in}}(\omega)-\sigma^{\dagger}_{\text{in}}(\omega)\right]}{\sqrt{2\kappa}(-i \omega+i\tilde{\omega}_{\text{m}}+\tilde{\gamma}_{\text{m}})}\\
&+G_e\left[\frac{\sqrt{2\kappa_{\text{c}}}a^{\dagger}_{\text{in}}(\omega)}{(-i\omega-i\Delta+\kappa)(-i\omega+i\tilde{\omega}_{\text{m}}+\tilde{\gamma}_{\text{m}})}-\frac{\sqrt{2\kappa_{\text{c}}}a_{\text{in}}(\omega)}{(-i\omega+i\Delta+\kappa)(-i\omega+i\tilde{\omega}_{\text{m}}+\tilde{\gamma}_{\text{m}})}\right]
\end{split}
\end{equation}
\end{widetext}
We have taken into account the optical spring effect, and the renormalized mechanical frequency and damping rate \cite{xuereb2011dissipative} are respectively $\tilde{\omega}_{\text{m}}=\omega_{\text{m}}+\omega_\text{o}$ and $\tilde{\gamma}_\text{m}=\gamma_{\text{m}}+\gamma_\text{o}$, where the $\omega_{\text{o}}$ and $\gamma_{\text{o}}$ are optical induced frequency and damping rate. When $\left|E\right|\gg\sqrt{2\kappa_\text{e}}\left|\sigma_{\text{e}}\right|$, we have
\begin{subequations}
\begin{align}
\omega_{\text{o}}=&(2\kappa_\text{e})^{3/2}\eta^2_\kappa\ \Im[\frac{E \sigma_{\text{e}}^{\ast}}{(-i\omega-i\Delta+\kappa)(-i\Delta+\kappa)}\nonumber\\
&\ \ \ \ \ \ \ \ \ \ \ \ \ \ \ \ \ -\frac{E^\ast \sigma_{\text{e}}}{(-i\omega+i\Delta+\kappa)(i\Delta+\kappa)}],\\
\gamma_{\text{o}}=& (2\kappa_\text{e})^{3/2}\eta^2_\kappa\ \Re[\frac{E \sigma_{\text{e}}^{\ast}}{(-i\omega-i\Delta+\kappa)(-i\Delta+\kappa)}\nonumber\\
&\ \ \ \ \ \ \ \ \ \ \ \ \ \ \ \ \ -\frac{E^\ast \sigma_{\text{e}}}{(-i\omega+i\Delta+\kappa)(i\Delta+\kappa)}].
\end{align}
\end{subequations}
Usually, we have $\omega_\text{m}\gg\omega_{\text{o}}$ and $\gamma_\text{m}\approx\gamma_{\text{o}}$ at the region we are interested in. The optical induced damping rate $\gamma_\text{o}$ is the major term to suppress the thermal phonon, and it also carries the information of force noise spectrum
and a numerical result is shown in Fig.~\ref{fig:damping}.
Unlike dispersive coupling, in our system, because of interference of the quantum noise, the optical induced damping reaches it's peak at blue detuning region so that we can realize dissipative optomechanical cooling in blue detuning.
But our system is also qualitatively different form the case of Michelson-Sagnac interferometer where the maximum of optical damping rate is always located at $\Delta=-0.5\omega_{\text{m}}$, however, in our system, the position of peak also depends on $\kappa_{\text{e}}$.
\\
\indent{}The steady state phonon number can be attained by calculating the noise spectrum and then transforming it back into time domain. We use the equation
$
\langle b^{\dagger}(t)b(t)\rangle=1/(2\pi)\int_{-\infty}^{\infty}d\omega S_{n_\text{m}}(\omega),
$
where $S_{n_\text{m}}(\omega)=\int_{-\infty}^{\infty} d\omega ' \langle b^{\dagger}(\omega)b(\omega ')\rangle$ is the noise spectrum of phonon.
A simple analytical result of steady state phonon number can be obtained in weak coupling regime,
\begin{equation}\label{phonon_number}
\begin{split}
\langle b^\dagger b\rangle_{\text{ss}}=&\frac{\gamma_{\text{m}}}{\tilde{\gamma}_\text{m}}n_{\text{th}}+\frac{G^2}{4\kappa_\text{e}\tilde{\gamma}_\text{m}}
+\frac{(\tilde{\gamma}_\text{m}+\kappa)\kappa_{\text{c}}{G^2_\text{e}}}{\tilde{\gamma}_\text{m} \kappa[(\tilde{\gamma}_\text{m}+\kappa)^2+(\Delta+\omega_\text{m})^2]}.
\end{split}
\end{equation}
The third term of this result is quite similar with \cite{xuereb2011dissipative}, and they only differs in the numerator of the optically induced phonon part, and this difference can be interpreted as a consequence of the extra quantum noise $\sigma_{\text{in}}$.
Our derivation has been based on linear response theory, the analytical solution could not describe high pump power case. But it is quite safe when the pump power is below the order of \text{mW}. The corresponding simulation is illustrated in Fig.~\ref{fig:phonon}.
The parameters we adopted are $D=30\ \mu\text{m}$, $\xi=0.01$, $A_{\text{eff}}/A=0.1$, $\omega_{\text{m}}=2\pi\times55\ \text{MHz}, \kappa_{\text{c}}=2\pi\times1\ \text{MHz},\kappa_\text{e}=2\pi\times45\ \text{MHz}, \eta_{\kappa}=2.2\times 10^{-3},$ $\gamma_{\text{m}}=2\pi\times10\ \text{Hz}$  \cite{chen2013graphene}.
In our setting, the thermal phonon number $n_{\text{th}}$ equals 100, corresponding to $0.26\ \text{K}$.
\begin{figure}
\centering{}
\includegraphics[width=0.5\textwidth]{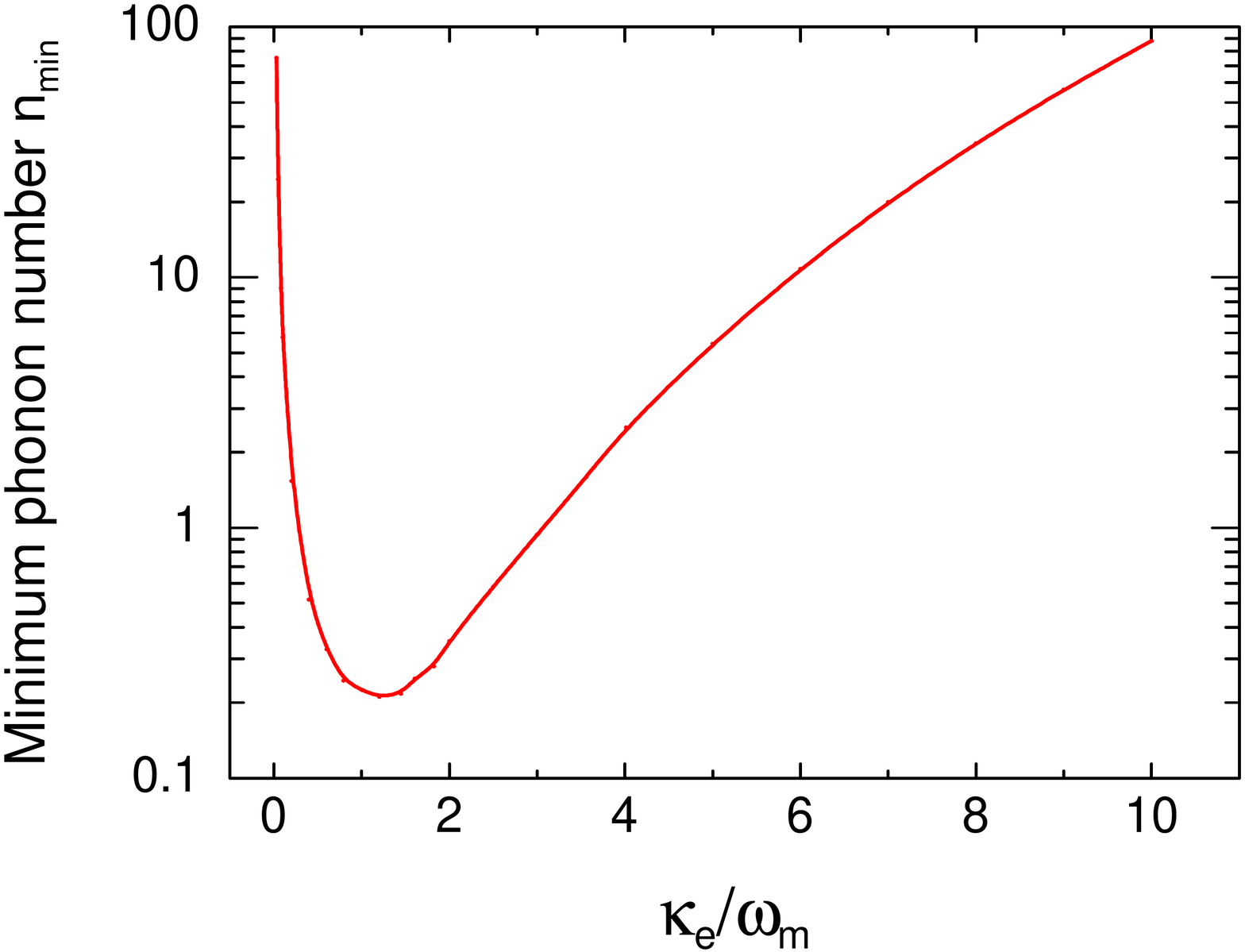}
\caption{\small (Color online)Minimum phonon number $n_{\text{min}}$ as a function of $\kappa_{\text{e}}$. we obtain this reliance by adjusting $\Delta$ and $P$ in Eq.~\ref{phonon_number} to obtain a local minimum for any given $\kappa_{\text{e}}$. All other parameters are fixed, and they are $\omega_{\text{m}}=2\pi\times 55 \ \text{MHz}, \kappa_{\text{c}}=2\pi\times1\ \text{MHz}, \eta_{\kappa}=2.2\times 10^{-3},$ $\gamma_{\text{m}}=\pi\times10\ \text{Hz}$, and the thermal phonon number $n_{\text{th}}=100$. }\label{fig:minimum_phonon}
\end{figure}
\\
\indent{}Further analysis shows that there exists an optimal cooling combination of $\Delta$ and pump power $P$ for a given absorptive damping rate $\kappa_{\text{e}}$ and ground state cooling can only be realized within a finite interval of $\kappa_\text{e}$.
The behavior of minimum phonon number $n_{\text{min}}(\kappa_{\text{e}})$ is illustrated in Fig.~\ref{fig:minimum_phonon}. As we can see, the optimal cooling requires $\kappa_{\text{e}}\thicksim\omega_{\text{m}}$ and ground state cooling can still be attained even when $\kappa_{\text{e}}$ approaches $4\omega_{\text{m}}$.\\
\section{Conclusion}
\label{sec:conclusion}
We have analyzed the dissipative optomechanical coupling between a single-layer graphene and an FP cavity in the membrane-in-the-middle setup with the inner degree of freedom of graphene properly handled. Unlike other dissipative coupling systems, we have verified the possibility of utilizing the absorptive decay to generate dissipative coupling.
Dissipative cooling of the graphene have been studied via the linearized quantum Langevin equations. For a single-layer graphene, ground state cooling can be achieved either in resolved-sideband regime or not.  In principle, our scheme could be generalized to any absorptive thin membrane.
This provides new insights for realizing dissipative optomechanical coupling within current experimental conditions.\\
\begin{acknowledgements}
We thank Wen-Ju Gu for helpful discussion of calculation details, and thank Fan Yang for his useful advices. The work was supported by the 973 program (2013CB328704), the NSFC (Grants No. 11004003, No. 11222440, and No. 11121091), and the Research Fund for the Doctoral Program of Higher Education (No. 20120001110068).
Lin-Da Xiao was supported by The Presents Fund for Undergraduate Research of Peking University.
Yu-Feng Shen was supported by The Beijing Undergraduate Innovational Experimentation Program.
\end{acknowledgements}

\bibliography{example1}
\end{document}